\documentclass[a4paper,12pt]{article}

\usepackage{amsmath}
\usepackage{amssymb}
\usepackage{amsthm}

\newcommand{\prt}{\partial}

\def\F{{\mathcal F}}

\def\ni{{\noindent}}

\def\be{\begin{equation}}
\def\ee{\end{equation}}

\newtheorem{theorem}{Theorem}

\newtheorem{lemma}{Lemma}

\theoremstyle{definition}

\begin{document}
\bibliographystyle{plain}

\title{{\Large\bf  Painlev\'e representation of Tracy-Widom$_\beta$ distribution for $\beta = 6$.}}
\author{Igor Rumanov \\  
{\small Dept. of Applied Mathematics, CU Boulder, Boulder, CO} \\
{\small e-mail: igor.rumanov@colorado.edu} }

\maketitle

\bigskip

\begin{abstract}
In \cite{betaFP1}, we found explicit Lax pairs for the soft edge of beta ensembles with even integer values of $\beta$. Using this general result, the case $\beta=6$ is further considered here. This is the smallest even $\beta$, when the corresponding Lax pair and its relation to Painlev\'e II (PII) have not been known before, unlike cases $\beta=2$ and $4$. It turns out that again everything can be expressed in terms of the Hastings-McLeod solution of PII. In particular, a second order nonlinear ordinary differential equation (ODE) for the logarithmic derivative of Tracy-Widom distribution for $\beta=6$ involving the PII function in the coefficients, is found, which allows one to compute asymptotics for the distribution function. The ODE is a consequence of a linear system of three ODEs for which the local singularity analysis yields series solutions with exponents in the set $4/3$, $1/3$ and $-2/3$.
\end{abstract}

\newpage

\section{Introduction and main result}

Beta ensembles of random matrices introduced by Dyson~\cite{DyBeta} were originally defined as Coulomb gas (fluid) of particles-eigenvalues for general values of Dyson index $\beta$ beyond the three most important cases $\beta=1, 2, 4$ known as real orthogonal (OE), complex unitary invariant (UE) and symplectic (SE) ensembles, respectively. The importance of general $\beta$ ensembles and the number of their applications grow fast in recent years due to the developments of Conformal Field Theory (CFT)~\cite{BPZ} connections with other subjects, see e.g.~\cite{AgEtAl11, BetaGarn} and references therein on the relation between $\beta$-ensembles and linear PDEs of~\cite{BPZ}. The applications include the AGT correspondence~\cite{AGT} relating CFT with supersymmetric quantum gauge field theories, and also condensed matter physics, e.g.~electronic transport in wires disordered by impurities and quantum Hall effect, see e.g.~\cite{F2010, DGIL94}. The tie of CFT itself with $\beta$-ensembles can be traced back to the times of its birth when the Coulomb gas representation of CFT correlation functions appeared in terms of Dotsenko-Fateev integrals~\cite{DF}. There are also the genuine matrix ensembles with general $\beta$ eigenvalue ditributions, first found in~\cite{DE02} for Gaussian and Laguerre weights and later extended to other measures, see~\cite{KRV13} and references therein. A comprehensive treatment of the available before the last several years results on $\beta$-ensembles and their applications is contained in~\cite{F2010}. 
\par The soft edge probability distributions (describing the largest eigenvalue when the matrix size $n\to \infty$) for $\beta=1, 2, 4$ have been known since the seminal works of Tracy and Widom~\cite{TW-Airy, TW-OrtSym} in terms of Hastings-Mcleod solution~\cite{HMcL80}\footnote{This solution in fact was found earlier by Ablowitz, Kruskal and Segur in~\cite{AbKrSeg}, based also on the results of~\cite{AbSegPIIlong}.} of Painlev\'e II. More recently, a one-parameter generalization of these distributions (describing e.g.~the soft edge limit of certain {\it spiked} ensembles or ensembles with external source) was shown to satisfy a diffusion-drift partial differential equation (PDE) for general values of $\beta$~\cite{EdSut, RRV, BV1}. For the three above special values its limit as the additional parameter $x$ tends to $+\infty$ is the corresponding Tracy-Widom distribution. However, the best available description up to date for the Tracy-Widom distributions of the other beta ensembles is the mentioned Fokker-Planck PDE, eq.~(1.1) below. 
\par This article is a sequel to~\cite{betaFP1}. Current results are a further demonstration of classical integrable structure present for values of $\beta$ beyond the three special ones where it was known or always expected. It should be somehow related to the quantum integrable structure of CFT with central charge $c\le 1$ found in~\cite{BLZ} but this is a matter of future investigation. We study the distribution function for the soft edge of (spiked) Dyson beta ensembles which satisfies the boundary value problem first considered by Bloemendal and Virag~\cite{BV1}:

$$
\left(\prt_t + \frac{2}{\beta}\prt_{xx} + (t-x^2)\prt_x\right)\F^{(\beta)}(t, x) = 0.   \eqno(1.1)
$$

\ni The boundary conditions ensure that the solution $\F^{(\beta)}$ to the Fokker-Planck (FP) eq.~(1.1) is a probability distribution function:

$$
\F^{(\beta)}(t,x) \to 0  \ \ \ \text{as } x \to -\infty, t < \infty,  \qquad   \F^{(\beta)}(t,x) \to 1 \quad \text{as } t, x \to +\infty \text{ together},
$$

$$
\F^{(\beta)}(t,x) \to F_{TW}^{(\beta)}(t)  \quad \text{as } x \to +\infty,  t \text{ finite}.  \eqno(1.2)
$$

\ni The last function $F_{TW}^{(\beta)}(t)$ is the Tracy-Widom distribution ($TW_\beta$). Equation (1.1) can be rightfully called quantum Painlev\'e II in imaginary time since besides the time derivative it contains operator which is the canonically quantized Painlev\'e II Hamiltonian with $2/\beta$ playing the role of Planck constant. All quantum Painlev\'e equations were introduced this way in~\cite{Nag11} and their special solutions as integrals over certain $\beta$-ensemble measures were found. In the $\beta$-ensembles of~\cite{Nag11} the parameter $\beta/2$ rather than $2/\beta$ as here corresponds to the Planck constant. This is the case for the averages of positive powers of characteristic polynomials w.r.t. $\beta$-eigenvalue (Coulomb gas) measures while our case corresponds to the ``dual"~\cite{Der09} ensembles with external source, see more details about this in~\cite{BetaGarn}. 
\par It will be convenient for us to consider the rescaled eq.~(1.1),

$$
\left(\kappa\prt_t + \prt_{xx} + (t-x^2)\prt_x\right)\F(t, x) = 0,   \eqno(1.3)
$$

\ni i.e.~eq.~(1.1) with $t$ and $x$ rescaled as $x\to x/\kappa^{1/3}$, $t\to t/\kappa^{2/3}$, $\kappa=\beta/2$. In \cite{betaFP1} we found explicit $2\times2$ matrix Lax pairs of the form

$$
\prt_x\left(\begin{array}{c}\F \\ G \end{array}\right) = L\left(\begin{array}{c}\F \\ G \end{array}\right),  \ \ \ \ \ \   \prt_t\left(\begin{array}{c}\F \\ G \end{array}\right) = B\left(\begin{array}{c}\F \\ G \end{array}\right),  \eqno(1.4)   
$$

\ni describing the soft edge 
of (spiked) random matrix beta ensembles, for {\it all even integer} Dyson indices $\beta$, such that $\F(t,x)$ solving eq.~(1.3) is the first component of their eigenvector. For positive integer $\kappa = \beta/2$ we obtained 

$$
L = \left(\begin{array}{cc} L_1 & L_+ \\ L_- & L_2 \end{array}\right) = \left(\begin{array}{cc} \frac{1}{2}(-v+L_d) & L_+ \\ -\frac{1}{2L_+}(\kappa B_d + \prt_xL_d + L_d^2/2 + f_v) & \frac{1}{2}(-v-L_d) \end{array}\right),   \eqno(1.5)  
$$

$$
B = \left(\begin{array}{cc} B_1 & B_+ \\ B_- & B_2 \end{array}\right) = \left(\begin{array}{cc} \frac{1}{2}\left(-x+\frac{U(t)+t^2/2}{\kappa}-\frac{\phi'}{\phi} + B_d\right) & -\frac{\prt_xL_+}{\kappa} \\ -\frac{2L_-\prt_xL_+ + \kappa \prt_tL_d - \kappa \prt_xB_d}{2\kappa L_+} & \frac{1}{2}\left(-x+\frac{U(t)+t^2/2}{\kappa}-\frac{\phi'}{\phi} - B_d\right)  \end{array}\right),   \eqno(1.6)
$$

\ni where 

$$
v= t-x^2,   \eqno(1.7)
$$

\ni is the drift function in the Fokker-Planck equation (1.3) (or (1.1)),

$$
L_+ = \phi(t)\prod_{n=1}^{\kappa}(x-Q_n(t)),   \eqno(1.8)    
$$


$$
L_d = -\frac{L_+}{\phi(t)}\cdot\sum_{n=1}^{\kappa}\frac{\kappa Q_n' - 2R_n}{(x-Q_n)\prod_{j\neq n}^{\kappa}(Q_n-Q_j)} = -\sum_{n=1}^{\kappa}(\kappa Q_n' - 2R_n)\prod_{j\neq n}^{\kappa}\frac{x-Q_j}{Q_n-Q_j},   \eqno(1.9)   
$$

$$
\kappa B_d = \kappa \phi'(t)/\phi + \sum_{n=1}^{\kappa}\frac{\kappa Q_n' - 2R_n}{x-Q_n}\left(\sum_{l=1}^{\kappa}\frac{\prod_{j\neq l}^{\kappa}(x-Q_j)}{\prod_{j\neq n}^{\kappa}(Q_n-Q_j)} - 1\right),   \eqno(1.10)    
$$






$$
f_v(t,x) = \kappa B_t - \prt_xv - v^2/2 = -\left(\frac{x^4}{2} - tx^2 + (\kappa-2)x - U(t) + \frac{\kappa \phi'(t)}{\phi}\right),  \eqno(1.11)   
$$

\ni $B_t\equiv Tr B=B_1+B_2$, $R_n = \sum_{j\neq n}^\kappa\frac{1}{Q_n - Q_j}$ and $U(t)$ is defined by

$$
\kappa U'(t) = -\sum_{n=1}^{\kappa}Q_n^2.   \eqno(1.12)
$$


\ni Function $\phi(t)$ remains arbitrary (but not identically zero), e.g.~one can take $\phi(t)\equiv 1$.
\par Functions $Q_n(t)$, $n=1, \dots, \kappa$, satisfy equations of motion for particles with Calogero interaction and additional time-dependent cubic external force which would lead to classical Painlev\'e II equation without the interaction~\cite{betaFP1}. Considered together with eq.~(1.12), they possess $\kappa$ explicit first integrals found in~\cite{betaFP1}, which we do not reproduce here because we find a different more convenient form of them in what follows.

\par The main result of the paper is

\begin{theorem}
The log-derivative of the rescaled Tracy-Widom distribution $\F_0(t)$ for $\beta=6$, where the Tracy-Widom distribution is $F_{TW}^{(\beta=6)}(t)=\F_0(\kappa^{2/3}t)=\F_0(3^{2/3}t)$, can be written as

$$
3(\ln\F_0)' = u - \frac{q^2}{u} + \eta,   \eqno(1.13)
$$

\ni where $q$ is the Hastings-McLeod solution of Painlev\'e II, $q''(t) = 2q^3+tq$, without free term, $u(t)=(q')^2-tq^2-q^4$ so that $u'(t)=-q^2$, and $\eta(t)$ satisfies the second order ODE:

$$
9\eta'' + 9\eta\eta' + \eta^3 - 4\left(3\left(\frac{q^2}{u}-u\right)' + t\right)\eta - 8\left(\frac{q^2}{u}-u\right)'' - 2 = 0.    \eqno(1.14)  
$$

\ni Equivalently, function $h_- = \eta - \frac{q^2}{u} = \frac{\nu}{\mu_-}$ can be found from the linear system of ODEs,

 $$
3q^2\mu_+' = (q^2)'\mu_+ - q^2\nu,    \eqno(1.15)  
$$

$$
3q^2\mu_-' = -(q^2)'\mu_- + q^2\nu,     \eqno(1.16)  
$$

$$
3q^2\nu' = 2q^4\mu_- - 2u\mu_+.     \eqno(1.17)  
$$

\end{theorem}

The local singularity analysis of system (1.15)--(1.17) shows that its exponents always belong to the set $\{4/3, 1/3, -2/3\}$ and there is always sufficient number of independent local series solutions. Having non-integer but rational exponents is in common with Garnier systems~\cite{GaussToP}, and in fact the functions $Q_n$ from eqs.~(1.8)--(1.12) have many properties of apparent singularities of Fuchsian ODEs making the linear problem for the Garnier systems, see~\cite{GaussToP, BetaGarn}. However, the cubes of our functions $\mu_{\pm}$ and $\nu$ are meromorphic in the complex plane as are their ratios, therefore the Painlev\'e property holds for them. The system (1.15)--(1.17) is also very special in the respect that while the leading exponents of its series solutions have integer differences, the series contain no logarithms.
\par The plan of the paper is the following. In section 2 we describe the Lax pair for $\kappa=3$ and demonstrate how its polynomiality leads to polynomial ODEs and their first integrals. We analyze the obtained system of equations in section 3 and find Painlev\'e II as well as the system (1.15)--(1.17) as its consequences. In section 4 we establish the relation (1.13), derive various equivalent forms of eq.~(1.14) and find the asymptotics of $(\ln\F_0)'$ as $t\to\pm\infty$. Section 5 presents the local singularity analysis of system (1.15)--(1.17). For completeness and comparison, in the Appendix we present the known~\cite{BV1, HBP3} simplest cases of $\kappa=1, 2$ (i.e.~$\beta = 2, 4$) which lead to Lax pairs for (classical) Painlev\'e II.

\section{The Lax pair for $\kappa=3$}

This is the first case beyond the previously known Lax pairs for Painlev\'e II, where the classical integrability has not been shown before. The general formulas of section 1 give:

$$
L_+ = \phi(t)(x-Q_1)(x-Q_2)(x-Q_3) = \phi(t)(x^3-e_1x^2+e_2x-e_3),   \eqno(2.1)  
$$

\ni where we introduced the elementary symmetric functions of $Q_k$, $e_j$ ($k, j = 1, 2, 3$). Then

$$
B_+ = -\frac{\prt_xL_+}{3} = -\phi(x^2 - 2e_1x/3 + e_2/3),   \eqno(2.2)  
$$

$$
B_t = -x + \frac{1}{3}\left(U+\frac{t^2}{2}\right) - \frac{\phi'}{\phi}.   \eqno(2.3)  
$$

\ni Since $L_d$ is now a quadratic polynomial in $x$ (see eq.~(1.9)), let

$$
L_d = q_2x^2-q_1x+q_0,  \eqno(2.4)  
$$

\ni where we could write each $q_j(t)$ in terms of $Q_k$-variables explicitly by eq.~(1.9) but we will not need this since, as we will see, $q_j$ are more convenient variables which will lead to polynomial first integrals of the system unlike the Garnier-like variables $Q_k$ in terms of which the first integrals are rational. Also, since $B_d$ is linear in $x$, see eq.~(1.10), let

$$
B_d - \frac{\phi'}{\phi} = d_1x - d_0.  \eqno(2.5)  
$$

\ni The next equation is a consequence of eqs.~(1.8)--(1.10), in fact, expressions (1.9) and (1.10) were chosen in order to satisfy it when $L_+$ was given by eq.~(1.8)~\cite{betaFP1}:

$$
L_+\cdot\kappa B_d + \prt_xL_+\cdot L_d = \kappa\prt_tL_+ + \prt_{xx}L_+   \eqno(2.6)  
$$

\ni Plugging eqs.~(2.1), (2.4) and (2.5) into eq.~(2.6) and equating the coefficients of the corresponding powers of $x$, we find $d_1$ and $d_2$ in terms of $q_j$ as well as first order ODEs for $e_j$:

$$
x^4: \qquad d_1 = -q_2,   \eqno(2.7)  
$$

$$
x^3: \qquad d_0 = \frac{e_1q_2}{3} - q_1,   \eqno(2.8)  
$$

$$
x^2: \qquad 3e_1' + (e_1^2-2e_2)q_2 - e_1q_1 + 3q_0 = 0,   \eqno(2.9)  
$$

$$
x^1: \qquad 3(e_2'+2) + (e_1e_2-3e_3)q_2 - 2e_2q_1 + 2e_1q_0 = 0,   \eqno(2.10)  
$$

$$
x^0: \qquad 3e_3' + 2e_1 + e_1e_3q_2 - 3e_3q_1 + e_2q_0 = 0.  \eqno(2.11)  
$$

\ni Now we determine $L_-$ from the corresponding component of general eq.~(1.5), using also eq.~(1.11),

$$
L_- = -\frac{L_d^2/2 + \prt_xL_d + 3B_d + f_v}{2L_+},   \eqno(2.12)  
$$

\ni Besides $L_-$, this equation turns out to yield the new polynomial first integrals. Explicitly the numerator of eq.~(2.12) reads:

$$
L_d^2/2 + \prt_xL_d + 3B_d + f_v = \frac{q_2^2-1}{2}x^4 - q_2q_1x^3 + \frac{2q_2q_0+q_1^2+2t}{2}x^2 - (q_1q_0+q_2+1)x + \frac{q_0^2}{2} + U - e_1q_2 + 2q_1
$$



$$
= (x^3-e_1x^2+e_2x-e_3)\left(\frac{q_2^2-1}{2}(x + e_1) - q_2q_1\right) + x^2I_2(t) + xI_1(t) + I_0(t).   \eqno(2.13) 
$$

\ni Clearing the denominator in eq.~(2.12) and matching powers of $x$ in the resulting equation implies that $L_-$ is a polynomial and so equals

$$
L_- = - \frac{(q_2^2-1)(x + e_1) - 2q_2q_1}{4\phi},   \eqno(2.14)  
$$

\ni and the remainder of the division of eq.~(2.13) by $L_+$ is zero which gives the three polynomial first integrals:

$$
I_2(t) \equiv  (e_1^2-e_2)\frac{q_2^2-1}{2} - e_1q_2q_1 + \frac{2q_2q_0+q_1^2+2t}{2} = 0,   \eqno(2.15)  
$$

$$
I_1(t) \equiv (e_3-e_1e_2)\frac{q_2^2-1}{2} + e_2q_2q_1 - q_1q_0 - q_2 - 1 = 0,   \eqno(2.16)  
$$

$$
I_0(t) \equiv e_1e_3\frac{q_2^2-1}{2} - e_3q_2q_1 + \frac{q_0^2}{2} + U - e_1q_2 + 2q_1 = 0.   \eqno(2.17)  
$$

\ni Now we determine the last entry, $B_-$, from the general formula, see eq.~(1.6), 

$$
3B_- = -\frac{2L_-\prt_xL_+ + 3\prt_tL_d - 3\prt_xB_d}{2L_+},   \eqno(2.18)  
$$

\ni and again the polynomiality of eq.~(2.18) multiplied by $L_-$ implies that 




$$
B_- = \frac{q_2^2-1}{4\phi}   \eqno(2.19)   
$$

\ni is given by the polynomial result of the division by $L_+$ and the remainder terms are equal to zero and thus give the three ODEs,

$$
3q_2' - 2e_1(q_2^2-1) + 3q_2q_1 = 0,   \eqno(2.20)  
$$

$$
3q_1' - (e_1^2+e_2)(q_2^2-1) + 2e_1q_2q_1 = 0, \eqno(2.21)   
$$ 

$$
3(q_0'+q_2) - \frac{(e_1e_2+3e_3)}{2}(q_2^2-1) + e_2q_2q_1 = 0.   \eqno(2.22)   
$$

\ni Adding to the system of ODEs (2.9)--(2.11) and (2.20)--(2.22) the ODE (1.12) for the function $U$, rewritten as

$$
3U' = -\sum_1^3Q_n^2 = 2e_2 - e_1^2,   \eqno(2.23)   
$$

\ni one can verify by tedious but straightforward calculation that

\begin{lemma}
Equations (2.15)--(2.17) are first integrals of the system of ODEs (2.9)--(2.11) and (2.20)--(2.23).
\end{lemma}

\ni Thus, three of the ODEs, e.g.~eqs.~(2.9)--(2.11), can be considered as redundant. The lemma also shows consistency of our Lax pair construction for $\kappa=3$ and gives the explicitly polynomial in $x$ expressions for Lax matrices (1.5) and (1.6):

$$
L = \left(\begin{array}{cc} L_1 & L_+ \\ L_- & L_2 \end{array}\right) = \left(\begin{array}{cc} \frac{x^2-t}{2} + \frac{q_2x^2-q_1x+q_0}{2} & \phi(t)(x^3-e_1x^2+e_2x-e_3) \\ -\frac{(q_2^2-1)(x + e_1) - 2q_2q_1}{4\phi} & \frac{x^2-t}{2} - \frac{q_2x^2-q_1x+q_0}{2} \end{array}\right),   \eqno(2.24)  
$$

$$
B = \left(\begin{array}{cc} B_1 & B_+ \\ B_- & B_2 \end{array}\right) = \left(\begin{array}{cc} \frac{-x + (U+t^2/2)/3 - q_2x - e_1q_2/3 + q_1}{2} & -\phi(t)(x^2 - 2e_1x/3 + e_2/3) \\ \frac{1}{4\phi}(q_2^2-1) & \frac{-x + (U+t^2/2)/3 + q_2x + e_1q_2/3 - q_1}{2} - \frac{\phi'}{\phi}  \end{array}\right).  \eqno(2.25)   
$$

\section{The $\kappa=3$ system.}

Continuing to analyze the system obtained for $\kappa=3$, we notice that the appearance of eqs.~(2.15)--(2.17) and (2.20)--(2.22) can be significantly simplified if one introduces a new function $r(t)$ such that

$$
e_1(q_2^2-1) = r(q_2^2-1) + 2q_2q_1.   \eqno(3.1)
$$

\ni Then the three first integrals eqs.~(2.15)--(2.17) can be written as, respectively,

$$
e_2(q_2^2-1) = re_1(q_2^2-1) + 2q_2q_0 + q_1^2 + 2t,   \eqno(3.2)
$$

$$
e_3(q_2^2-1) = re_2(q_2^2-1) + 2q_1q_0 + 2 + 2q_2,   \eqno(3.3)
$$

$$
0 = re_3(q_2^2-1) + q_0^2 + 2U - 2e_1q_2 + 4q_1.   \eqno(3.4)
$$

\ni We also rewrite eqs.~(2.20)--(2.22) as

$$
6q_2' = (e_1 + 3r)(q_2^2-1),  \eqno(3.5)
$$

$$
3q_1' = (e_2 + re_1)(q_2^2-1),  \eqno(3.6)
$$

$$
6q_0' = (3e_3 + re_2)(q_2^2-1) - 6q_2.  \eqno(3.7)
$$

\ni We observe that eqs.~(3.1)--(3.4) have a nice ``telescopic" structure and their linear combination $r^3\cdot(3.1) + r^2\cdot(3.2) + r\cdot(3.3) + (3.4)$ gives

$$
(r^2q_2 + rq_1 + q_0)^2 - r^4 + 2tr^2 + 2r + 2U +2[(r-e_1)q_2 + 2q_1] = 0.   \eqno(3.8)
$$

\ni Anticipating what follows we introduce the new function $u_r$ to replace $U$,

$$
u_r = U + (r-e_1)q_2 + 2q_1 = U + \frac{r-e_1}{q_2} = U - \frac{2q_1}{q_2^2-1},   \eqno(3.9)
$$

\ni where we used eq.~(3.1) in the last two equalities. Thus, we can consider the new first integral

$$
(r^2q_2 + rq_1 + q_0)^2 - r^4 + 2tr^2 + 2r + 2u_r = 0   \eqno(3.10)
$$

\ni as replacing eq.~(3.4). Then we have to derive the ODE for $u_r$ replacing eq.~(2.23). On the one hand, using the second last expression in eq.~(3.9), we write 

$$
3u_r' = 3U' + \frac{3r'-3e_1'}{q_2} - \frac{(r-e_1)\cdot3q_2'}{q_2^2},
$$

\ni and, using eqs.~(2.23), (2.9), (3.5) and (3.1), after some cancellations we obtain






$$
u_r' = \frac{r' + rq_1 + q_0}{q_2}.   \eqno(3.11)
$$

\ni On the other hand, from the last expression on the right-hand side of eq.~(3.9), we find, with the help of eqs.~(2.23), (3.5), (3.1) and (3.6),

$$
3u_r' = 3U' + \frac{2q_2q_1\cdot6q_2'}{(q_2^2-1)^2} - \frac{2\cdot3q_1'}{q_2^2-1}  = 2e_2-e_1^2 + (e_1-r)(e_1+3r) - 2(e_2+re_1),
$$




\ni i.e. 

$$
u_r' = -r^2,   \eqno(3.12)
$$

\ni justifying the introduction of $u_r$. Then eq.~(3.11) implies 

$$
r' + r^2q_2 + rq_1 + q_0 = 0,   \eqno(3.13)
$$

\ni and eq.~(3.10) now means that

$$
(r')^2 - r^4 + 2tr^2 + 2r + 2u_r = 0,   \eqno(3.14)
$$

\ni which, together with eq.~(3.12), yields Painlev\'e II equation for the function $r$,

$$
r'' = 2r^3 - 2tr - 1.   \eqno(3.15)
$$

\ni This is incidentally the same Painlev\'e II which is satisfied by the function $Q(t) = -q'(t)/q$ for $\kappa=1$, see eq.~(A7) of the Appendix. So we identify  

$$
r = -\frac{q'}{q},  \qquad q''=2q^3+tq,  \eqno(3.16)
$$

\ni where $q$ is the Hastings-McLeod solution of Painlev\'e II.
\par Now we can eliminate $q_0$ expressing it from eq.~(3.13) and substituting into the other equations. Then we are left with only two independent ODEs to resolve, with coefficients depending on the known function $r$. It is convenient to choose eqs.~(3.5) and (3.6) as such and, after using eqs.~(3.1) and (3.2) (besides eq.~(3.13)) to eliminate $e_1$ and $e_2$, they become, respectively,

$$
3q_2' = 2r(q_2^2-1) + q_2q_1,  \eqno(3.17)
$$

$$
3q_1' = 2rq_2q_1 + q_1^2 + 2(t-r^2) - 2r'q_2.  \eqno(3.18)
$$

\begin{lemma} 
Equations (3.17), (3.18) are equivalent to the linear system (1.15)--(1.17) of the main theorem.
\end{lemma}

{\it Proof:} We notice a combination $r_1 = 2rq_2 + q_1$ appearing in both the last equations, differentiating it we get

$$
3r_1' = r_1^2 + 4r'q_2 + 2(t-3r^2),   \eqno(3.19)
$$

\ni and introducing a new function $\chi$ such that

$$
r_1 \equiv 2rq_2 + q_1 = -3\frac{\chi'}{\chi},  \eqno(3.20)
$$

\ni we rewrite eq.~(3.17) as



$$
3(q_2\chi)' = -2r\chi.  \eqno(3.21)
$$

\ni In turn, eq.~(3.18) becomes



$$
3(q_1\chi)' = -2[r'q_2\chi + (r^2-t)\chi]   \eqno(3.22)
$$

\ni after substituting eq.~(3.20). We introduce now two new functions by

$$
q_2 = \frac{\mu}{\chi}, \qquad q_1 = \frac{\nu}{\chi},   \eqno(3.23)
$$

\ni which allows us to get a system of three linear equations equivalent to eqs.~(3.17), (3.18):

$$
3\chi' = -2r\mu - \nu,   \eqno(3.24)
$$

$$
3\mu' = -2r\chi,   \eqno(3.25)
$$

$$
3\nu' = -2(r'\mu + (r^2-t)\chi).  \eqno(3.26)
$$

\ni (Eq.~(3.19) is redundant being a consequence of them.) At last we express everything in terms of Painlev\'e transcendent $q$ instead of $r$. We use eqs.~(3.16) written as

$$
q' = -rq, \qquad r' = r^2 - t - 2q^2,   \eqno(3.27)
$$

\ni and introduce function $u$ such that

$$
u = (q')^2 - q^4 - tq^2,  \qquad u' = -q^2.   \eqno(3.28)
$$

\ni This function is well known as a Hamiltonian function of Painlev'e II, see e.g.~\cite{GaussToP}. Then 

$$
r' + r^2 - t = 2(r^2 - t - q^2) = 2\frac{u}{q^2}.   \eqno(3.29)
$$

\ni Using the above and introducing also

$$
\mu_{\pm} = \mu \pm \chi   \eqno(3.30)
$$





\ni transforms eqs.~(3.24)--(3.26) into the system (1.15)--(1.17). $\square$

\section{Tracy-Widom distribution for $\kappa=3$ and auxiliary functions}

The linear system (1.15)--(1.17) with coefficients depending on $q^2$ and $u$ completely characterizes the $\kappa=3$ ($\beta=6$) case since all the important functions can be readily found from $\mu_+$, $\mu_-$ and $\nu$ as we will show now. Return to the Quantum Painlev\'e II -- the Fokker-Planck equation (1.3). Let us consider the asymptotic expansion of $\F(t,x)$ as $x\to\infty$,

$$
\F(t,x) = \sum_{n=0}^\infty\frac{\F_n(t)}{x^n},   \eqno(4.1)
$$

\ni which agrees with the boundary conditions eq.~(1.2) corresponding to the sought solution $\F(t,x)$ being a probability distribution function. The function $\F_0(t)$ is the rescaled Tracy-Widom-beta ($TW_{\beta}$ in short) distribution, $\F_0(t)=F_{TW}^{\beta}(t/\kappa^{2/3})$, recall going to eq.~(1.3) from eq.~(1.1). Substituting eq.~(4.1) into eq.~(1.3), one finds recursion relations for the expansion coefficients, i.e.,~since 

$$
\prt_t\F = \sum_{n=0}^\infty\frac{\F_n'(t)}{x^n}, \quad \prt_x\F = -\sum_{n=2}^\infty\frac{(n-1)\F_{n-1}(t)}{x^n}, \quad \prt_{xx}\F = \sum_{n=3}^\infty\frac{(n-1)(n-2)\F_{n-2}(t)}{x^n}, \eqno(4.2)
$$

\ni one obtains

$$
\F_1 = -\kappa\F_0', \qquad \F_2 = -\frac{\kappa}{2}\F_1' = \frac{\kappa^2}{2}\F_0'',  \qquad \F_3 = \frac{t\F_1-\kappa\F_2'}{3} = -\frac{\kappa^3\F_0'''+2t\kappa\F_0'}{6},  \eqno(4.3)
$$

\ni and

$$
(n+1)\F_{n+1} = -\kappa\F_n' + (n-1)t\F_{n-1} - (n-1)(n-2)\F_{n-2}, \qquad n\ge 3.   \eqno(4.4)
$$

\ni Thus, all the functions $\F_n(t)$ can be recursively found in terms of $\F_0(t)$ and its derivatives. Due to the Lax pair (1.4), $\F(t,x)$ also satisfies~\cite{betaFP1} a first order ODE,

$$
\kappa\prt_t\F + P(t,x)\prt_x\F + b(t,x)\F = 0,  \eqno(4.5)
$$

\ni with $P(t,x)$ and $b(t,x)$ explicitly known for integer $\kappa$ in terms of the entries of the Lax matrices (1.5), (1.6):

$$
P(t,x) = -\kappa\frac{B_+}{L_+} = \sum_{n=1}^\kappa\frac{1}{x-Q_n(t)},   \eqno(4.6)
$$


$$
b(t,x) = \frac{1}{2}\sum_{n=1}^\kappa\frac{\kappa Q_n' + t - Q_n^2 - 2R_n}{x-Q_n} - \frac{1}{2}\left(\frac{t^2}{2} + U(t) + \sum_{n=1}^\kappa Q_n\right),   \eqno(4.7)
$$

\ni where

$$
R_n = \sum_{j\neq n}^\kappa\frac{1}{Q_n - Q_j}.   \eqno(4.8)
$$





\ni One can expand eq.~(4.5) at large $x$ as well for every $\kappa$ and we will explore the full consequences of this elsewhere. For our current purposes we need only the first terms of this expansion, the limit of eq.~(4.5) as $x\to\infty$, which yields

$$
\kappa\F_0' - \frac{1}{2}\left(\frac{t^2}{2} + U(t) + e_1\right)\F_0 = 0,   \eqno(4.9)   
$$

\ni where we used that, for integer $\kappa$,

$$
\sum_{n=1}^\kappa Q_n = e_1.  \eqno(4.10)   
$$

{\it Remark.} Eq.~(4.9), however, holds for every $\kappa$, integer or not, with $e_1$ and $U$ defined from expansion of $P(t,x)$ and $b(t,x)$ at large $x$ which is valid and has the same form for all $\kappa$, unlike eq.~(4.10). 
\par For the case $\kappa=3$ at hand, eq.~(4.9) says:

$$
3(\ln\F_0)' = \frac{1}{2}\left(\frac{t^2}{2} + U(t) + e_1\right),  \eqno(4.11)  
$$

\ni which implies a simple connection of $\F_0$ with functions considered in the previous sections. First, from eqs.~(3.1) and (3.23) we have

$$
e_1 = r + \frac{2q_2q_1}{q_2^2-1} = r + \frac{2\mu\nu}{\mu^2-\chi^2},  \eqno(4.12)   
$$

\ni and, using eqs.~(3.27) and (3.30),

$$
e_1 = -\frac{q'}{q} + \frac{\nu}{\mu_+} + \frac{\nu}{\mu_-} \equiv -\frac{g'}{2g} + \frac{\nu}{\mu_+} + \frac{\nu}{\mu_-}.  \eqno(4.13)  
$$

\ni Here and further on we denote $g=q^2$. Next, from eqs.~(3.9), (3.23) and (3.30) we find

$$
U = u_r + \frac{2q_1}{q_2^2-1} = u_r + \frac{2\chi\nu}{\mu^2-\chi^2} = u_r + \frac{\nu}{\mu_-} - \frac{\nu}{\mu_+}.  \eqno(4.14)   
$$

\ni At last, using eqs.~(3.14), (3.27) and (3.28), we express $u_r$ as




$$
u_r = -\frac{(r')^2}{2} + \frac{r^4}{2} - tr^2 - r =  2u + \frac{q'}{q} - \frac{t^2}{2} = 2u + \frac{g'}{2g} - \frac{t^2}{2}.  \eqno(4.15)   
$$

\ni Substituting eqs.~(4.13)--(4.15) into eq.~(4.11) we finally obtain

$$
3(\ln\F_0)' = u + \frac{\nu}{\mu_-}.  \eqno(4.16)   
$$

\ni In the rest of this section, we are going to derive various forms of the ODE (1.14) of the main theorem and then find the asymptotics of $\Phi(t)\equiv3(\ln\F_0)'(t)$. 

\subsection{Various forms of the ODE (1.14)}

For further convenience, let us denote

$$
h_-(t) = \frac{\nu}{\mu_-}, \qquad h_+(t) = \frac{\nu}{\mu_+}.  \eqno(4.17)    
$$

\ni Consider again system (1.15)--(1.17). Eq.~(1.17) can be rewritten in two ways, using eq.~(4.17), 







$$
3g(h_+\mu_+)' = 2g^2\mu_- - 2u\mu_+,  \eqno(4.18)   
$$

$$
3g(h_-\mu_-)' = 2g^2\mu_- - 2u\mu_+.  \eqno(4.19)   
$$ 

\ni Dividing eq.~(4.18) by $\mu_+$ and eq.~(4.19) by $\mu_-$, and using eqs.~(1.15), (1.16) and definitions (4.17), yields equations for $h_+$ and $h_-$, respectively,

$$
3gh_+' + g'h_+ - gh_+^2 + 2u = 2g^2\frac{h_+}{h_-},   \eqno(4.20)  
$$

$$
3gh_-' - g'h_- + gh_-^2 - 2g^2 = -2u\frac{h_-}{h_+}.   \eqno(4.21)  
$$

\ni Expressing $h_+$ from eq.~(4.21) and substituting into eq.~(4.20) yields a second order ODE for $h_-$ with coefficients depending on $g$, $g'$ and $u$,

$$
\frac{2uh_- - 3(3gh_-' - g'h_- + gh_-^2 - 2g^2)'}{3gh_-' - g'h_- + gh_-^2 - 2g^2} - h_- + 2\frac{g'}{g} - 3\frac{g}{u} = 0.   \eqno(4.22)  
$$

\ni Using eqs.~(3.28) rewritten as

$$
u = \frac{(g')^2}{4g} - g^2 - tg, \qquad u' = -g,  \eqno(4.23)  
$$

\ni and their consequence

$$
g'' = 6g^2 + 4tg + 2u = \frac{(g')^2}{g} + 2g^2 - 2u,  \eqno(4.24)  
$$

\ni eq.~(4.22) can be brought to the form



$$
9h_-'' + 9\left(h_- + \frac{g}{u}\right)h_-' + h_-^3 + 3\frac{g}{u}h_-^2 - \left(3\frac{g'}{u} + 12g + 4t\right)h_- - 8g' - \frac{6g^2}{u} = 0.   \eqno(4.25)  
$$

\ni Similarly, expressing $h_-$ from eq.~(4.20) and substituting into eq.~(4.21) gives a second order ODE for $h_+$, which, after using eqs.~(4.23) and (4.24), finally becomes



$$
9g^2h_+'' - 3g^2h_+h_+' - g^2h_+^3 + 2gg'h_+^2 - [(g')^2 - 6g^3 +4ug]h_+ - 6g^2 - 8ug' = 0.  \eqno(4.26)  
$$

\ni (We record it for completeness although it is not used further on). Using eq.~(4.16), one can also derive the corresponding ODE for $\Phi \equiv 3(\ln\F_0)' = h_- + u$ from eq.~(4.25):

$$
9\Phi'' + 9\left(\Phi + \frac{g}{u} - u\right)\Phi' + \Phi^3 + 3\left(\frac{g}{u}-u\right)\Phi^2 + 
$$

$$
+ \left(3u^2 - 3\frac{g'}{u} - 9g - 4t\right)\Phi + 4g' + \frac{3g^2}{u} + 4tu + 6ug - u^3 = 0.   \eqno(4.27)  
$$

\ni It is more convenient for finding asymptotics of $\Phi$ as $t\to+\infty$.
\par The simplest form of the final equation is reached, however, if one uses $\eta = h_- + g/u$ as the dependent variable. Then eq.~(4.25) acquires the form

$$
9\eta'' + 9\eta\eta' + \eta^3 - 4(3\Gamma' + t)\eta - 8\Gamma'' - 2 = 0,   \eqno(4.28)  
$$

\ni (which is the equation (1.14) of the main theorem) where 

$$
\eta = h_- + \frac{g}{u},  \qquad \Phi = \eta - \Gamma,  \qquad \Gamma = \frac{g}{u} - u.   \eqno(4.29)  
$$

\ni To derive it we used that, by eqs.~(4.23) and (4.24),

$$
\left(\frac{g}{u}\right)' = \frac{g'}{u} + \left(\frac{g}{u}\right)^2,   \qquad   \left(\frac{g}{u}\right)'' = 3\frac{gg'}{u^2} + 2\left(\frac{g}{u}\right)^3 + 6\frac{g^2}{u} + 4t\frac{g}{u} + 2.   \eqno(4.30)  
$$

\ni However, eq.~(4.28) turns out to be not convenient for finding the asymptotics of $\Phi$ as $t\to\pm\infty$.

\subsection{Asymptotics of $\Phi(t)$ as $t\to+\infty$}  

When $t\to +\infty$ it is best to use eq.~(4.27). One can verify that $\Phi$ has the following asymptotic expansion:

$$
\Phi = \frac{e^{-4t^{3/2}/3}}{t^4}\sum_{n=0}^\infty\frac{\phi_n}{t^{3n/2}}   \qquad  \text{as } t \to +\infty.   \eqno(4.31)  
$$

\ni Due to the exponential factor eq.~(4.27) linearizes in this limit (as is the case for the involved Painlev\'e II itself) and takes form (after multiplying by $u$)

$$
9u\Phi'' + 9g\Phi' - (3g' + 4tu)\Phi + 4ug' + 3g^2 + 4tu^2 = 0,   \eqno(4.32)  
$$

\ni which can be solved by the series (4.31). The Painlev\'e functions expand in this limit as

$$
q = \frac{e^{-2t^{3/2}/3}}{t^{1/4}}\sum_{n=0}^\infty\frac{C_n}{t^{3n/2}},  \qquad  g = \frac{e^{-4t^{3/2}/3}}{t^{1/2}}\sum_{n=0}^\infty\frac{g_n}{t^{3n/2}},  \qquad  u = \frac{e^{-4t^{3/2}/3}}{t}\sum_{n=0}^\infty\frac{u_n}{t^{3n/2}},   \eqno(4.33)  
$$

\ni where the coefficients are related by $g_0 = 2u_0$, 

$$
n\ge 1: g_n = 2u_n + \frac{3n-1}{2}u_{n-1},  \quad  g_n = \sum_{l=0}^nC_lC_{n-l}, \ C_{n+1} = -\frac{(1+6n)(5+6n)}{48(n+1)}C_n,  \eqno(4.34)  
$$

\ni and $C_0=\frac{1}{2\sqrt\pi}$ is known e.g.~from~\cite{FIKN}. Then we get

$$
g' = e^{-4t^{3/2}/3}\sum_{n=0}^\infty\frac{(g')_n}{t^{3n/2}},  \qquad  g^2 = \frac{e^{-8t^{3/2}/3}}{t}\sum_{n=0}^\infty\frac{(g^2)_n}{t^{3n/2}},  \eqno(4.35)  
$$

\ni where

$$
(g')_0 = -2g_0, n\ge1: (g')_n = -(2g_n + (3n/2 - 1)g_{n-1}),   \qquad  (g^2)_n = \sum_{l=0}^ng_lg_{n-l},    \eqno(4.36)  
$$





\ni Substituting everything into eq.~(4.32) one verifies that the first two orders in powers of $t$ cancel identically, i.e.

$$
4(ug')_0 + 3(g^2)_0 + 4(u^2)_0 = 0,  \qquad   4(ug')_1 + 3(g^2)_1 + 4(u^2)_1 = 0,   \eqno(4.37)  
$$

\ni (since $g_0=2u_0$, $g_1=2u_1+u_0$, $(g')_0=-2g_0$ etc.) and the others recursively determine coefficients $\phi_n$, $n\ge 0$:




$$
(32u_0-3(g')_0-18g_0)\phi_n=4g_0\phi_n = -(4ug'+3g^2+4u^2)_{n+2} - 
$$

$$
- \sum_{l=0}^{n-1}[32u_{n-l} - 3(g')_{n-l} - 18g_{n-l} + 27(3+2l)u_{n-1-l} - 9(5+3l)g_{n-1-l}/2]\phi_l + 
$$

$$
+ 9\sum_{l=0}^{n-2}(1+3l/2)(2+3l/2)u_{n-2-l}\phi_l = 0, \quad n\ge0.   \eqno(4.38)  
$$

\ni Thus one obtains e.g.


$$
4g_0\phi_0 = -4(u_0(g')_2+u_1(g')_1+u_2(g')_0) - 3(2g_0g_2+g_1^2) - 8(2u_0u_2+u_1^2),   \eqno(4.39)  
$$


\ni and, using the above relations between coefficients, $\phi_0 = (3g_1+g_0)/8=(6C_0C_1+C_0^2)/8 = 3C_0^2/64=3/(256\pi)$. Upon dividing by $3$ and rescaling back $t\to 3^{2/3}t$ this matches predictions from~\cite{ForAsymBeta, BorNad}.

\subsection{Asymptotics of $\Phi(t)$ as $t\to-\infty$}  

It is convenient to use eq.~(4.25) for the function $h_-$ here, we multiply it by $u$ to clear denominators. The expansions for Painlev\'e functions in this limit are

$$
q = \sum_{n=0}^\infty C_n\left(-\frac{t}{2}\right)^{-3n+1/2},  \quad  g = \sum_{n=0}^\infty g_n\left(-\frac{t}{2}\right)^{-3n+1},  \quad  u = \sum_{n=0}^\infty u_n\left(-\frac{t}{2}\right)^{-3n+2},   \eqno(4.40)  
$$

\ni where the coefficients are related by

$$
g_n = \left(1 - \frac{3n}{2}\right)u_n,  \qquad  g_n = \sum_{l=0}^nC_lC_{n-l},
$$

\begin{displaymath}  
4C_n = \frac{36(n-1)^2-1}{16}C_{n-1} - \sum_{k+l+m=n; k,l,m\ge 0}^{k,l,m\le n-1}2C_kC_lC_m, \qquad C_0 = 1.  \eqno(4.41)  
\end{displaymath}  

\ni Then

$$
g' = -\frac{1}{2}\sum_{n=0}^\infty\frac{(g')_n}{(-t/2)^{3n}},  \qquad  g^2 = \frac{t^2}{4}\sum_{n=0}^\infty\frac{(g^2)_n}{(-t/2)^{3n}},  \qquad  ug' = -\frac{t^2}{8}\sum_{n=0}^\infty\frac{(ug')_n}{(-t/2)^{3n}},   \eqno(4.42)  
$$

\ni where

$$
(g')_n = \frac{(3n-1)(3n-2)}{2}u_n,  \qquad  (g^2)_n = \sum_{l=0}^ng_lg_{n-l},   \qquad   (ug')_n = \sum_{l=0}^nu_l(g)'_{n-l}.   \eqno(4.43)  
$$




\ni Substituting the series solutions of the form $h_- = \tilde h_0(-t)^\alpha + \dots$, where $\alpha$ is the leading exponent, into eq.~(4.25), one finds that there are two possibilities, $\alpha=1/2$ and $\alpha=-1$. To describe the Tracy-Widom distribution, one has to pick $\alpha=1/2$ (unlike we did in the first version of the paper) to match the results obtained by other methods, see below\footnote{We are very grateful to Peter Forrester for pointing out the discrepancies to the author immediately after the first version of the paper appeared online.}. Then eq.~(4.25) gives $\tilde h_0^2 = 2$, and again there are two choices and the right one is $\tilde h_0 = -\sqrt2$. This leads to the solution series such that

$$
h_- = -(-2t)^{1/2}\sum_{n=0}^\infty \frac{h_n}{(-t/2)^{3n/2}}, \qquad  h_-' = \frac{1}{2}\sum_{n=0}^\infty \frac{(h')_n}{(-t/2)^{(3n+1)/2}}, \qquad  (h')_n = (1-3n)h_n,   \eqno(4.44)  
$$

$$
h_-'' = \frac{1}{8}\sum_{n=0}^\infty \frac{(h'')_n}{(-t/2)^{3(n+1)/2}}, \qquad  (h'')_n = (1-9n^2)h_n.   \eqno(4.45)   
$$

\ni Substituting everything into eq.~(4.25) one finds $-u_0h_0^3 + 3u_0g_0h_0 - 2u_0h_0 = 0$ for $n=0$ and, since $u_0=g_0=1$, one gets $h_0^2=1$ and chooses now $h_0=1$. Then also $(h')_0=(h'')_0=h_0=1$ by eqs.~(4.44), (4.45). The general recursion relation for the coefficients is





$$
8[-(uh_-^3)_n + 3(ugh_-)_n - 2(uh)_n] + 4[-\frac{9}{4}(uh_-h_-')_{n-1} + 3(gh_-^2)_{n-1} + (g'u)_{(n-1)/2} - \frac{3}{2}(g^2)_{(n-1)/2}] +
$$

$$
+ \frac{9}{8}(uh_-'')_{n-2} + \frac{9}{2}(gh_-')_{n-2} - 3(g'h_-)_{n-2} = 0,   \eqno(4.46)  
$$

\ni where it is implied that

$$
(fh_-^k)_n = \sum_{j=0}^nf_{(n-j)/2}(h_-^k)_j, \quad  (h_-^k)_j=\sum_{m_1+\dots+m_k=j}h_{m_1}\dots h_{m_k}, \quad  f = u, g, g', ug,
$$

$$
(uh_-h_-')_n = \sum_{j=0}^nu_{(n-j)/2}\sum_{k=0}^jh_k(h')_{j-k},  \quad  (uh_-'')_n = \sum_{j=0}^nu_{(n-j)/2}(h'')_j,  \quad  (gh_-')_n = \sum_{j=0}^ng_{(n-j)/2}(h')_j,
$$

\ni and all the quantities with half-integer indices are zero. E.g.~for $n=1$ we have





$$
8[-u_0(h_-^3)_1 + 3u_0g_0h_1 - 2u_0h_1]  + 4[-\frac{9}{4}u_0h_0(h')_0 + 3g_0h_0^2 + (g')_0u_0 - \frac{3}{2}g_0^2] = 0,   \eqno(4.47)  
$$

\ni and, using that $(h'')_0=(h')_0=h_0=u_0=g_0=(g')_0=1$, $(h_-^3)_1=3h_0^2h_1=3h_1$, we obtain $h_1=1/16$. Thus, the first terms of the expansion for $\Phi$ are






$$
\Phi = u + h_- = \frac{t^2}{4} - \frac{1}{8t} + \dots - (-2t)^{1/2} - (-2t)^{1/2}\frac{1}{16(-t/2)^{3/2}} + \dots = 
$$

$$
= \frac{t^2}{4} - (-2t)^{1/2} + \frac{1}{8t} + \dots,   \eqno(4.48)  
$$

\ni and, after taking into account that $(\ln\F_0)' = \Phi/3$ and rescaling back $t\to 3^{2/3}t$, see the main theorem in the first section, this matches the known results~\cite{BEMN}, see also formula (2.16) in~\cite{ForRev12},

$$
\ln F_{TW}^{\beta} = -\beta\frac{|t|^3}{24} + \frac{\sqrt2 (\beta/2-1)}{3}|t|^{3/2} + \frac{\beta/2 + 2/\beta - 3}{8}\ln |t| + \dots = -\frac{|t|^3}{4} + \frac{2\sqrt2}{3}|t|^{3/2} + \frac{1}{24}\ln |t| + \dots
$$

\section{Local singularity analysis of $\kappa=3$ system.}

It is interesting and illuminating to verify if the system (1.15)--(1.17) satisfies the Painlev\'e property, i.e.~if its solutions are single-valued. The author would like to thank M.~Ablowitz for the suggestion to do it. 
\par As is well-known, see e.g.~\cite{AbCl, FIKN}, the only singularities of all solutions of Painlev\'e II in the complex plane are simple poles, and all poles of the Hastings-McLeod solution $q$ lie in two symmetric sectors of angle $\pi/3$ around imaginary axis with the vertex at the origin, see e.g.~\cite{Nov12} for the clear statement of this result; it should be noted that the singularity sectors and global asymptotics of Painlev\'e II were described much earlier e.g.~in~\cite{Kap92}. Every solution of Painlev\'e II $q$ always has a Laurent expansion around a point $t=t_0$,

$$
q = z^l\sum_{n=0}^\infty a_nz^n,  \eqno(5.1)
$$

\ni where $z=t-t_0$. All its poles and zeros are simple, see e.g.~\cite{AbCl, FIKN}, so the exponent $l$ can be $-1$, $0$ or $1$. Consider the corresponding expansions for the functions entering the coefficients of eqs.~(1.15)--(1.17). As follows from eq.~(5.1),

$$
g \equiv q^2 = z^{2l}\sum_{n=0}^\infty g_nz^n,  \qquad   g' = z^{2l-1}\sum_{n=0}^\infty (2l+n)g_nz^n,    \eqno(5.2)
$$



\ni and, using also that $u'=-q^2=-g$, we find that 

$$
u = u_0 - z^{2l+1}\sum_{n=0}^\infty \frac{g_n}{2l+1+n}z^n,  \eqno(5.3)  
$$

\ni Then we consider expansions

$$
\mu_+ = z^{m_+}\sum_{n=0}^\infty K_nz^n, \qquad \mu_- = z^{m_-}\sum_{n=0}^\infty M_nz^n, \qquad \nu = z^{m_{\nu}}\sum_{n=0}^\infty S_nz^n. \eqno(5.4)  
$$



\ni Substituting all the expansions into the eqs.~(1.15)--(1.17), we obtain, respectively,

$$
3z^{m_+-1}\sum_{n=0}^\infty z^n \sum_{j=0}^ng_{n-j}(j+m_+)K_j = z^{m_+-1}\sum_{n=0}^\infty z^n \sum_{j=0}^n(2l+n-j)g_{n-j}K_j - z^{m_{\nu}}\sum_{n=0}^\infty z^n\sum_{j=0}^ng_{n-j}S_j,   \eqno(5.5)  
$$

$$
3z^{m_--1}\sum_{n=0}^\infty z^n \sum_{j=0}^ng_{n-j}(j+m_-)M_j = -z^{m_--1}\sum_{n=0}^\infty z^n \sum_{j=0}^n(2l+n-j)g_{n-j}M_j + z^{m_{\nu}}\sum_{n=0}^\infty z^n\sum_{j=0}^ng_{n-j}S_j,   \eqno(5.6)  
$$

$$
3z^{m_{\nu}+2l-1}\sum_{n=0}^\infty z^n \sum_{j=0}^ng_{n-j}(j+m_{\nu})S_j = 2z^{m_-+4l}\sum_{n=0}^\infty z^n \sum_{j=0}^n(g^2)_{n-j}M_j - 
$$

$$
- 2u_0z^{m_+}\sum_{n=0}^\infty K_nz^n + 2z^{m_+ + 2l+1}\sum_{n=0}^\infty z^n\sum_{j=0}^n\frac{g_{n-j}}{2l+1+n-j}K_j,   \eqno(5.7)  
$$

\ni As follows from eqs.~(5.5), (5.6), $m_{\nu}\ge m_+ -1$ and $m_{\nu}\ge m_- -1$ in general. It is convenient to proceed from here considering separately the cases when $q$ has pole ($l=-1$) and $q$ has zero ($l=1$). As for the case of a regular point of $q$ ($l=0$), the solutions of linear ODEs are always regular at the regular points of their coefficients.

\subsection{Local behavior near a pole of $q$.}


\begin{theorem}
Near every pole of $q$, there are three types of solutions of the system (1.15)--(1.17): 1)exponents $m_+=m_-=4/3$, $m_{\nu}=1/3$ and one free constant; 2)exponents $m_+=m_-=1/3$, $m_{\nu}=-2/3$ and two free constants; 3)exponents $m_+=m_{\nu}=-2/3$, $m_-=1/3$ and three free constants. The third type is thus generic.
\end{theorem}

We always can combine these linearly independent solutions locally around each simple pole of a function $q$, the solution of Painlev\'e II without constant term. \\

{\it Proof:} At a simple pole of $q$, one has $a_0=\pm1$, $a_1=0$, therefore $g_0=a_0^2=1$ and $g_1 = 2a_0a_1 = 0$. Now eqs.~(5.5)--(5.7) read, respectively,

$$
z^{m_{\nu}}\sum_{n=0}^\infty z^n\sum_{j=0}^ng_{n-j}S_j = z^{m_+-1}\sum_{n=0}^\infty z^n \sum_{j=0}^ng_{n-j}(-3m_+-2+n-4j)K_j,   \eqno(5.8)  
$$

$$
z^{m_{\nu}}\sum_{n=0}^\infty z^n\sum_{j=0}^ng_{n-j}S_j = z^{m_--1}\sum_{n=0}^\infty z^n \sum_{j=0}^ng_{n-j}(3m_- -2 + n + 2j)M_j.   \eqno(5.9)  
$$

$$
3z^{m_{\nu}-1}\sum_{n=0}^\infty z^n \sum_{j=0}^ng_{n-j}(j+m_{\nu})S_j = 2z^{m_- - 2}\sum_{n=0}^\infty z^n \sum_{j=0}^n(g^2)_{n-j}M_j - 
$$

$$
- 2u_0z^{m_+ + 2}\sum_{n=0}^\infty K_nz^n + 2z^{m_+ +1}\sum_{n=0}^\infty z^n\sum_{j=0}^n\frac{g_{n-j}}{n-j-1}K_j.   \eqno(5.10)  
$$

\ni Analyzing their first terms, one can conclude that there are two different possibilities for the values of the exponents: either $m_+=m_-=m$, $m_{\nu}=m-1$, or $m_+=m_{\nu}=-2/3$, $m_-=1/3$. In both cases $m_{\nu}=m_- - 1$ and eq.~(5.9) gives coefficients $S_n$ in terms of $M_n$ recursively (we use the facts that $g_0=1$ and $g_1=0$):

$$
S_n = (3(n+m_-)-2)M_n + \sum_{j=0}^{n-2}g_{n-j}[(n-2+3m_-+2j)M_j - S_j],  \eqno(5.11)  
$$

\ni which is valid for all $n\ge 0$ if the terms with negative indices or sum with upper limit less than the lower are understood as absent. 
\par {\it Case $m_+=m_-=m$, $m_{\nu}=m-1$.} \\
We substitute $S_n$ from eq.~(5.11) into eq.~(5.10) and obtain the recursion relation which determines the coefficients $M_n$,


$$
9(n+m-4/3)(n+m-1/3)M_n = \sum_{j=0}^{n-2}g_{n-j}[3(n-j)S_j + (4 - 3(n+m-1)(3(n+m)-2-2(n-j)))M_j]
$$

$$
- 2K_{n-3} + 2\sum_{j=0}^{n-4}\left[(g^2)_{n-4-j}M_j + \frac{g_{n-3-j}}{n-4-j}K_l\right].   \eqno(5.12)  
$$

\ni Putting $n=0$ in eq.~(5.12), since the first coefficient $M_0$ is non-zero by definition, the possible exponents $m$ are found to be $m=4/3$ or $m=1/3$. Putting $n=1$ instead gives

$$
(m-1/3)(m+2/3)M_1 = 0,
$$

\ni which means that either $m=1/3$ and $M_1$ remains undetermined or, if $m=4/3$, then $M_1=0$. The difference of eqs.~(5.8) and (5.11) yields the recursive expression for coefficients $K_n$, see below.
\par {\it Case $m=4/3$}: only one constant $M_0$ is free, $M_1=0$ (which entails also $S_1=K_1=0$) and the other coefficients are recursively determined by

$$
S_n = (3n+2)M_n + \sum_{j=0}^{n-2}g_{n-j}[(n+2+2j)M_j - S_j].  \eqno(5.13)  
$$

$$
3(n+2)K_n = -(3n+2)M_n - \sum_{j=0}^{n-2}g_{n-j}[(n+2+2j)M_j - (n-6-4j)K_j],  \eqno(5.14)  
$$

$$
9n(n+1)M_n = \sum_{j=0}^{n-2}g_{n-j}[3(n-j)S_j + (4 - (3n+1)(n+2+2j))M_j] - 2K_{n-3} +
$$

$$
+ 2\sum_{j=0}^{n-4}\left[(g^2)_{n-4-j}M_j + \frac{g_{n-3-j}}{n-4-j}K_j\right],   \eqno(5.15)  
$$

\ni where eq.~(5.13) follows from eq.~(5.11), eq.~(5.14) -- from the difference of eq.~(5.8) and eq.~(5.11), and eq.~(5.15) -- from eq.~(5.12).
\par {\it Case $m=1/3$}: two constants, $M_0$ and $M_1$, are free and the other coefficients are recursively determined in the same way as above,

$$
S_n = (3n-1)M_n + \sum_{j=0}^{n-2}g_{n-j}[(n-1+2j)M_j - S_j],  \eqno(5.16)  
$$

$$
3(n+1)K_n = -(3n-1)M_n - \sum_{j=0}^{n-2}g_{n-j}[(n-1+2j)M_j - (n-3-4j)K_j],  \eqno(5.17)  
$$

$$
9n(n-1)M_n = \sum_{j=0}^{n-2}g_{n-j}[3(n-j)S_j + (4 - (3n-2)(n-1+2j))M_j] - 2K_{n-3} +
$$

$$
+ 2\sum_{j=0}^{n-4}\left[(g^2)_{n-4-j}M_j + \frac{g_{n-3-j}}{n-4-j}K_j\right].   \eqno(5.18)  
$$

\par {\it Case $m_+=m_{\nu}=-2/3$, $m_-=1/3$.}  \\
Then in eqs.~(5.8), (5.10) the constant $K_0$ remains free (undetermined). Eq.~(5.11) yields  

$$
S_n = (3n-1)M_n + \sum_{j=0}^{n-1}g_{n-j}[(n+2j-1)M_j - S_j].   \eqno(5.19)  
$$

\ni The difference of eqs.~(5.8) and (5.11) gives recursion ($n\ge0$)

$$
3(n+1)K_{n+1} = \sum_{j=0}^n[(n-4j+1)g_{n+1-j}K_j - (n+2j-1)g_{n-j}M_j]  \eqno(5.20)  
$$

\ni (e.g.~$3K_1=M_0$). At last, eq.~(5.10) leads to 

$$
(3n-2)S_n = 2M_n + \sum_{j=0}^{n-1}[2(g^2)_{n-j}M_j - (3j-2)g_{n-j}S_j] - 2u_0K_{n-3} + 2\sum_{j=0}^{n-2}\frac{g_{n-2-j}}{n-3-j}K_j,   \eqno(5.21)  
$$

\ni valid for $n\ge 0$ in the same sense as before. The first two relations in eq.~(5.21), i.e.~for $n=0,1$, are the same as in eq.~(5.19), while the third, when compared to its $n=2$ case, relates $M_2$ and $K_0$ by 

$$
K_0 = -3(3M_2+g_2M_0).   \eqno(5.22)  
$$

\ni The coefficients $M_n$ for $n\ge 3$ are recursively determined from substituting eq.~(5.19) into eq.~(5.21),

$$
9n(n-1)M_n = \sum_{j=0}^{n-1}[(2(g^2)_{n-j} - (3n-2)(n+2j-1)g_{n-j})M_j - 3(n-j)g_{n-j}S_j] - 
$$

$$
- 2u_0K_{n-3} + 2\sum_{j=0}^{n-2}\frac{g_{n-2-j}}{n-3-j}K_j.   \eqno(5.23)  
$$

\ni Thus, we obtain the series solution with three free constants, e.g.~$M_0, M_1$ and $K_0$. $\square$

\subsection{Local behavior near a zero of $q$.}

\begin{theorem}
Near every zero of $q$, there are three types of solutions of the system (1.15)--(1.17): 1)exponents $m_+=m_-=4/3$, $m_{\nu}=1/3$ and one free constant; 2)exponents $m_+=m_-=1/3$, $m_{\nu}=-2/3$ and two free constants; 3)exponents $m_+=1/3$, $m_-=m_{\nu}=-2/3$ and three free constants. The third type is thus generic.
\end{theorem}

Again we can always consider their linear combination with the needed three arbitrary constants to choose. \\
{\it Proof:} Now $l=1$ and from Painlev\'e II and eq.~(3.28) we find that 

$$
u_0 = g_0,   \qquad   g_1=a_1=0.     \eqno(5.24)  
$$

\ni In place of eqs.~(5.8)--(5.10) we have

$$
z^{m_{\nu}}\sum_{n=0}^\infty z^n\sum_{j=0}^ng_{n-j}S_j = z^{m_+-1}\sum_{n=0}^\infty z^n \sum_{j=0}^ng_{n-j}(2-3m_+ +n-4j)K_j,   \eqno(5.25)  
$$

$$
z^{m_{\nu}}\sum_{n=0}^\infty z^n\sum_{j=0}^ng_{n-j}S_j = z^{m_--1}\sum_{n=0}^\infty z^n \sum_{j=0}^ng_{n-j}(3m_- +2 + n + 2j)M_j.   \eqno(5.26)  
$$

$$
3z^{m_{\nu}+1}\sum_{n=0}^\infty z^n \sum_{j=0}^ng_{n-j}(j+m_{\nu})S_j = 2z^{m_- + 4}\sum_{n=0}^\infty z^n \sum_{j=0}^n(g^2)_{n-j}M_j - 
$$

$$
- 2u_0z^{m_+}\sum_{n=0}^\infty K_nz^n + 2z^{m_+ +3}\sum_{n=0}^\infty z^n\sum_{j=0}^n\frac{g_{n-j}}{n-j+3}K_j.   \eqno(5.27)  
$$

\ni One also has two cases here to consider separately, either $m_+=m_-=m$, $m_{\nu}=m-1$ as the first case for poles, or $m_+=1/3$, $m_{\nu}=m_-=-2/3$.

\par {\it Case $m_+=m_-=m$, $m_{\nu}=m-1$.} \\
\ni Eq.~(5.25) determines $S_n$ recursively,

$$
g_0S_n = g_0(2-3(n+m))K_n + \sum_{j=0}^{n-1}g_{n-j}[(2-3m+n-4j)K_j-S_j],  \eqno(5.28)  
$$

\ni and, similarly, the difference of eqs.~(5.25) and (5.26) determines $M_n$,   

$$
g_0(3(n+m)+2)M_n = g_0(2-3(n+m))K_n + \sum_{j=0}^{n-1}g_{n-j}[(2-3m+n-4j)K_j-(2+3m+n+2j)M_j].  \eqno(5.29)  
$$

\ni At last, eq.~(5.27), after substituting eq.~(5.28), results in the recursion for $K_n$,



$$
[3g_0(n+m-1)(2-3(n+m)) + 2u_0]K_n = 3\sum_{j=0}^{n-1}g_{n-j}[(n-j)S_j - (n+m-1)(2-3m+n-4j)K_j] + 
$$

$$
+ 2\sum_{j=0}^{n-3}\frac{g_{n-3-j}}{n-j}K_j + 2\sum_{j=0}^{n-4}(g^2)_{n-4-j}M_j,  \eqno(5.30)  
$$

\ni Again, putting $n=0$ in eq.~(5.30) implies (since $K_0\neq 0$) that the possible exponents are $m=4/3$ or $m=1/3$. But now, using eq.~(5.24), the $n=1$ component of recursion (5.30) reads:

$$
3g_0(m-1/3)(m+2/3)K_1 = -g_1[S_0 + 3m(m-1)K_0] = 0, 
$$

\ni so either $m=4/3$ and $K_1=0$ or $m=1/3$ and $K_1$ remains undetermined. If $m=4/3$, then eq.~(5.30) becomes 

$$
9g_0n(n+1)K_n = -3\sum_{j=0}^{n-1}g_{n-j}[(n-j)S_j + (n+1/3)(4j+2-n)K_j] + 
$$

$$
- 2\sum_{j=0}^{n-3}\frac{g_{n-3-j}}{n-j}K_j - 2\sum_{j=0}^{n-4}(g^2)_{n-4-j}M_j.   \eqno(5.31)  
$$

\ni If $m=1/3$, then, after using also eq.~(5.24), eq.~(5.30) turns into

$$
-9g_0n(n-1)K_n = 3\sum_{j=0}^{n-1}g_{n-j}[(n-j)S_j - (n+m-1)(2-3m+n-4j)K_j] + 
$$

$$
+ 2\sum_{j=0}^{n-3}\frac{g_{n-3-j}}{n-j}K_j + 2\sum_{j=0}^{n-4}(g^2)_{n-4-j}M_j,  \eqno(5.32)  
$$

\ni Thus, the solution with $m=4/3$ has one free constant $K_0$ and the solution with $m=1/3$ has two free constants $K_0$ and $K_1$. 
\par {\it Case $m_+=1/3$, $m_{\nu}=m_-=-2/3$.}
In the same way as for the other cases, eqs.~(5.25)--(5.27) give the following recursion relations:

$$
g_0S_n = g_0(1-3n)K_n + \sum_{j=0}^{n-1}g_{n-j}[(1+n-4j)K_j-S_j],  \eqno(5.33)  
$$

$$
3g_0nM_n = \sum_{j=0}^{n-1}g_{n-j}[(1+n-4j)K_j-(n+2j)M_j],  \eqno(5.34)  
$$

\ni for $n\ge1$ with $M_0$ undetermined, and

$$
9g_0n(n-1)K_n = \sum_{j=0}^{n-1}g_{n-j}[(3n-2)(1+n-4j)K_j - 3(n-j)S_j] - 
$$

$$
- 2\sum_{j=0}^{n-3}\frac{g_{n-3-j}}{n-j}K_j - 2\sum_{j=0}^{n-4}(g^2)_{n-4-j}M_j.    \eqno(5.35)  
$$

\ni Again, since $g_1=0$, $K_0$ and $K_1$ remain undetermined and here we have another solution with three free constants $M_0$, $K_0$ and $K_1$. $\square$
\par In conclusion, we have a very special linear system (1.15)--(1.17), the solutions of which have leading exponents which differ by integers but there are no logarithms. We always have local series solutions with three free constants. While the exponents are non-integer, the cubes and ratios of our functions $\mu_{\pm}$ and $\nu$ are meromorphic in the complex plane, therefore the Painlev\'e property persists for the equations involving only these combinations, e.g.~various forms of eq.~(1.14) or eq.~(4.26).
\par A nonlinear integrable ODE without the Painlev\'e property appeared recently in~\cite{DubKap} where it described isomonodromic deformation dynamics with respect to a parameter in equation $P_I^2$, the second equation in Painlev\'e I hierarchy, a fourth order ODE which universally appears under scaling around generic gradient catastrophe points of hyperbolic nonlinear PDEs. It seems that such examples are just tips of a large array of integrable systems without Painlev\'e property yet to be identified.    

\bigskip

{\bf\large Acknowledgments.} It is a pleasure to thank M.~Ablowitz, R.~Halburd and R.~Maier for useful discussions, P.~Forrester for prompt message which helped notice and correct mistakes in the first posted version of the paper, and the referees for careful reading and suggestions to improve the presentation. Partial support by NSF grants DMS-0645756 and DMS-0905779 is gratefully acknowledged.

\section*{Appendix: Lax pairs for $\kappa=1, 2$}

The Lax pairs obtained here must be gauge equivalent to the originally derived Lax representations of Painlev\'e II~\cite{FN80, JM81} but the question of explicit transformations between different such Lax pairs is outside the scope of this paper.

\subsection*{$\kappa = 1$.}

Then, according to the above formulas of section 1, we have

$$
L_+ = \phi(t)(x-Q(t)),  \qquad  B_+ = -\frac{\prt_xL_+}{\kappa} = -\phi(t),   \eqno(A1)  
$$


$$
B_t = -\int_0^x\prt_tv dx + b_t(t) = -x + b_t(t), \qquad b_t(t) = \frac{t^2}{2} + U(t) - \frac{\phi'(t)}{\phi},   \eqno(A2)   
$$

$$
L_d = -Q'(t),  \qquad  B_d = \frac{\phi'(t)}{\phi},   \eqno(A3)   
$$


\ni and therefore

$$
L_1 = -\frac{v+Q'}{2} = \frac{x^2-t}{2} - \frac{Q'}{2},  \qquad L_2 = -\frac{v-Q'}{2} = \frac{x^2-t}{2} + \frac{Q'}{2},   \eqno(A4)  
$$

$$
B_1 = -\frac{x}{2} + \frac{1}{2}\left(\frac{t^2}{2} + U\right),  \qquad B_2 = -\frac{x}{2} + \frac{1}{2}\left(\frac{t^2}{2} + U\right) - \frac{\phi'(t)}{\phi}.   \eqno(A5)  
$$

\ni Substituting the expression for $L_-$, we obtain

$$
L_- = -\frac{\kappa B_d + \prt_xL_d + L_d^2/2 + V(v)}{2L_+} = \frac{x^4 - 2tx^2 - 2x - (Q')^2 - 2U}{4\phi(x-Q)} = 
$$

$$
= \frac{x^3 + Qx^2 + (Q^2-2t)x + Q(Q^2-2t) - 2}{4\phi} + \frac{Q^2(Q^2-2t) - 2Q - (Q')^2 - 2U}{4\phi(x-Q)},
$$

\ni however, the only nonpolynomial term with would-be pole is in fact equal to zero: equations

$$
U = \frac{Q^4}{2} - tQ^2 - Q - \frac{(Q')^2}{2}, \qquad U' = -Q^2,    \eqno(A6)  
$$

\ni give the right Painlev\'e II equation for $Q$:

$$
Q'' = 2Q^3 - 2tQ - 1,   \eqno(A7)  
$$

\ni which is satisfied by $Q = -q'/q$, $q$ being the Hastings-McLeod solution of Painlev\'e II with free parameter zero:

$$
q'' = 2q^3 + tq.   \eqno(A8)  
$$

\ni As we will see, this cancellation of terms which would make $L_-$ have poles in $x$ is the simplest example of the general phenomenon for the above Lax pairs constructed in~\cite{betaFP1}. Thus,

$$
L_- = \frac{x^3 + Qx^2 + (Q^2-2t)x + Q(Q^2-2t) - 2}{4\phi},   \eqno(A9)  
$$

\ni and now, substituting the explicit expression for $B_-$, we meet another example of similar cancellation of polar terms:

$$
B_- = -\frac{2\prt_xL_+L_- + \kappa \prt_tL_d - \kappa \prt_xB_d}{2L_+} = -\frac{x^4 - 2tx^2 - 4Q(Q^2-t)x + Q^2(3Q^2-2t)}{4\phi(x-Q)^2},   
$$

\ni but $x^4 - 2tx^2 - 4Q(Q^2-t)x + Q^2(3Q^2-2t) = (x-Q)^2(x^2 + 2Qx + 3Q^2 - 2t)$, so finally

$$
B_- = -\frac{x^2 + 2Qx + 3Q^2 - 2t}{4\phi}.   \eqno(A10)  
$$

\ni It is convenient to express everything in terms of $q$ instead of $Q = -q'/q$, and to choose the arbitrary function $\phi(t)$ as $\phi = -q$, then

$$
L_+ = -qx - q', \qquad B_+ = q, \qquad L_d = -\left(\frac{q'}{q}\right)^2 + 2q^2 + t, \qquad B_d = \frac{q'}{q},   \eqno(A11)  
$$


\ni and, introducing function $u(t)$ as in eq.~(3.28) we get 



$$
U + \frac{t^2}{2} = 2u + \frac{q'}{q}.   \eqno(A12)  
$$

\ni Finally we obtain the Lax pair for $\kappa=1$ in terms of Hastings-McLeod $q(t)$:

$$
L = \left(\begin{array}{cc} L_1 & L_+ \\ L_- & L_2 \end{array}\right) = \left(\begin{array}{cc} \frac{x^2-t}{2} + \frac{1}{2}\left(-\left(\frac{q'}{q}\right)^2 + 2q^2 + t\right) & -qx - q' \\ -\frac{x^3 - q'x^2/q + ((q'/q)^2-2t)x - q'((q'/q)^2-2t)/q - 2}{4q} & \frac{x^2-t}{2} - \frac{1}{2}\left(-\left(\frac{q'}{q}\right)^2 + 2q^2 + t\right) \end{array}\right),   
$$

$$
B = \left(\begin{array}{cc} B_1 & B_+ \\ B_- & B_2 \end{array}\right) = \left(\begin{array}{cc} \frac{1}{2}(-x + 2u + \frac{q'}{q}) & q \\ \frac{x^2 - 2q'x/q + 3(q'/q)^2 - 2t}{4q} & \frac{1}{2}(-x+ 2u - \frac{q'}{q})  \end{array}\right).  
$$

\ni We note that it is different from the Baik-Rains~\cite{BR01} pair appeared in this context in~\cite{BBP, BV1}:

$$
\prt_{t}\left(\begin{array}{c}f \\ g \end{array}\right) = \left(\begin{array}{cc}  0 & q \\ q & -x  \end{array}\right)\left(\begin{array}{c}f \\ g \end{array}\right),    
$$

$$
\prt_{x}\left(\begin{array}{c}f \\ g \end{array}\right) = \left(\begin{array}{cc}  q^2 & -qx-q' \\ -qx+q' & x^2 - t - q^2  \end{array}\right)\left(\begin{array}{c}f \\ g \end{array}\right).   
$$


\subsection*{$\kappa = 2$.}

Here we have $Q_1(t) = - Q_2(t) = Q(t)$ ($Q$ is now different from the previous section, see below). From general formulas of section 1 we get

$$
L_+ = \phi(t)(x-Q_1)(x-Q_2) = \phi(x^2-Q^2),  \qquad  B_+ = -\frac{\prt_xL_+}{2} = -\phi(t)x,   \eqno(A13)  
$$


$$
B_t = -\int_0^x\prt_tv dx + b_t(t) = -x + b_t(t), \qquad b_t(t) = \frac{1}{2}\left(\frac{t^2}{2} + U(t)\right) - \frac{\phi'(t)}{\phi},   \eqno(A14)  
$$

$$
L_d = -\left(2\frac{Q'}{Q} - \frac{1}{Q^2}\right)x,   \qquad   B_d = \frac{\phi'(t)}{\phi} + 2\frac{Q'}{Q} - \frac{1}{Q^2},   \eqno(A15)   
$$


\ni and therefore

$$
L_1 = \frac{x^2-t}{2} - \left(2\frac{Q'}{Q} - \frac{1}{Q^2}\right)\frac{x}{2},  \qquad L_2 = \frac{x^2-t}{2} + \left(2\frac{Q'}{Q} - \frac{1}{Q^2}\right)\frac{x}{2},   \eqno(A16)  
$$

$$
B_1 = -\frac{x}{2} + \frac{1}{4}\left(\frac{t^2}{2} + U\right) + \frac{1}{2}\left(2\frac{Q'}{Q} - \frac{1}{Q^2}\right),  \ \ B_2 = -\frac{x}{2} + \frac{1}{4}\left(\frac{t^2}{2} + U\right) - \frac{1}{2}\left(2\frac{Q'}{Q} - \frac{1}{Q^2}\right) - \frac{\phi'(t)}{\phi}.   \eqno(A17)  
$$

$$
f_v = -\frac{x^4}{2} + tx^2 + U - \frac{2\phi'}{\phi},   \eqno(A18)  
$$

\ni Substituting the expression for $L_-$ from eq.~(1.5), we obtain


$$
L_- = \frac{x^4/2 - \left(2(Q' - 1/2Q)^2 + tQ^2\right)x^2/Q^2 - U - 2Q'/Q + 1/Q^2}{2\phi(x^2-Q^2)} = 
$$

$$
= \frac{x^2 + Q^2 - 2t - (2Q'/Q-1/Q^2)^2}{4\phi} + \frac{Q^2(Q^2-2t) - 4(Q')^2 + 1/Q^2- 2U}{4\phi(x^2-Q^2)},
$$

\ni but again the last fraction is equal to zero: equations

$$
U = \frac{(Q^2)^2}{2} - tQ^2 - \frac{((Q^2)')^2}{2Q^2} + \frac{1}{2Q^2}, \qquad U' = -Q^2,    \eqno(A19)  
$$

\ni lead to the following Painlev\'e II equation for $Q^2$:

$$
2Q^2(Q^2)'' - ((Q^2)')^2 = 2(Q^2)^2(Q^2 - t) - 1,   \eqno(A20)  
$$

\ni which is satisfied by $Q^2 = 2q^2 + 2q' + t$ with the same $q$ satisfying eq.~(A8). Thus

$$
L_- = \frac{x^2 + Q^2 - 2t - (2Q'/Q-1/Q^2)^2}{4\phi},   \eqno(A21)  
$$

\ni and now, substituting the explicit expression for $B_-$ from eq.~(1.6), 


$$
2B_- = \frac{-x\left(x^2 + Q^2 - 2t - (2Q'/Q-1/Q^2)^2\right)/2 + x(2Q'/Q-1/Q^2)'}{\phi(x^2-Q^2)},   
$$

\ni and, using eqs.~(A19) and (A20), this simplifies to just

$$
B_- = -\frac{x}{4\phi}.   \eqno(A22)   
$$

\ni It is now convenient to choose the arbitrary function $\phi(t)$ so that 

$$
\phi'/\phi = -q,
$$

\ni then express everything in terms of $q$ instead of $Q$, $Q^2 = 2q^2 + 2q' + t$. To this end, we record

$$
(Q^2)' = 2qQ^2 + 1, \qquad \Longrightarrow \qquad 2\frac{Q'}{Q} - \frac{1}{Q^2} = 2q,
$$

$$
L_+ = \phi(t)(x^2 - 2q^2 - 2q' - t), \qquad B_+ = -\phi(t)x, 
$$

$$
L_d = -2qx, \qquad B_d = q, \qquad L_- = \frac{x^2 + 2q' - t - 2q^2}{4\phi}  \eqno(A23)  
$$


\ni and, again introducing function $u(t)$ given by eq.~(3.28), we get

$$
U + \frac{t^2}{2} = 2(u - q),  \qquad  B_t = -x + u.    \eqno(A24)  
$$

  
\ni Finally we obtain the Lax pair for $\kappa=2$ in terms of Hastings-McLeod $q(t)$, which is exactly the pair we derived in~\cite{HBP3} by the hard-to-soft edge limit transition from Lax pair for quantum Painlev\'e III~\cite{Nag11} describing the hard edge for beta ensembles~\cite{RR08, RRZ}:

$$
L = \left(\begin{array}{cc} L_1 & L_+ \\ L_- & L_2 \end{array}\right) = \left(\begin{array}{cc} \frac{x^2-t}{2} - qx & \phi(t)(x^2 - 2q^2 - 2q' - t) \\ \frac{x^2 + 2q' - t - 2q^2}{4\phi} & \frac{x^2-t}{2} + qx \end{array}\right),   
$$

$$
B = \left(\begin{array}{cc} B_1 & B_+ \\ B_- & B_2 \end{array}\right) = \left(\begin{array}{cc} \frac{1}{2}(-x + u + q) & -\phi(t)x \\ -\frac{x}{4\phi} & \frac{1}{2}(-x + u - q)  \end{array}\right).  
$$

























\end{document}